\documentclass[preprint]{aastex}

\lefthead{Itoh et al.}
\righthead{}

\begin{document}

\title{A Young Brown Dwarf Companion to DH~Tauri\footnote{Based on data collected at the Subaru Telescope, which is operated by the National Astronomical Observatory of Japan.}}
\author{Yoichi Itoh\altaffilmark{2},
Masahiko Hayashi\altaffilmark{3},
Motohide Tamura\altaffilmark{4},
Takashi Tsuji\altaffilmark{5},
Yumiko Oasa\altaffilmark{2},
Misato Fukagawa\altaffilmark{6},
Saeko S. Hayashi\altaffilmark{3},
Takahiro Naoi\altaffilmark{6}, 
Miki Ishii\altaffilmark{3},
Satoshi Mayama\altaffilmark{7}, 
Jun-ichi Morino\altaffilmark{3},  
Takuya Yamashita\altaffilmark{3},
Tae-Soo Pyo\altaffilmark{3}, 
Takayuki Nishikawa\altaffilmark{8},  
Tomonori Usuda\altaffilmark{3},
Koji Murakawa\altaffilmark{3}, 
Hiroshi Suto\altaffilmark{3}, 
Shin Oya\altaffilmark{3}, 
Naruhisa Takato\altaffilmark{3}, 
Hiroyasu Ando\altaffilmark{4}, 
Shoken M. Miyama\altaffilmark{4}, 
Naoto Kobayashi\altaffilmark{5}, 
and
Norio Kaifu\altaffilmark{4},
}
\altaffiltext{2}{Graduate School of Science and Technology, Kobe University,
1-1 Rokkodai, Nada, Kobe, Hyogo 657-8501, Japan, 
yitoh@kobe-u.ac.jp}
\altaffiltext{3}{Subaru Telescope,
650 N'Aohoku Pl., Hilo, Hawaii 96720, USA}
\altaffiltext{4}{National Astronomical Observatory,
2-21-1 Osawa, Mitaka, Tokyo 181-8588, Japan}
\altaffiltext{5}{Institute of Astronomy, The University of Tokyo,
2-21-1 Osawa, Mitaka, Tokyo 181-8588, Japan}
\altaffiltext{6}{Graduate School of Science, The University of Tokyo,
2-21-1 Osawa, Mitaka, Tokyo 181-8588, Japan}
\altaffiltext{7}{Graduate School of Science and Engineering,
Waseda University, 3-4-1 Okubo, Shinjuku, Tokyo 169-8555, Japan}
\altaffiltext{8}{Department of Astronomical Science, Graduate University 
for Advanced Studies (Sokendai),
650 N. A'ohoku Place, Hilo, Hawaii 96720, USA}

\begin{abstract}
We present the detection of a young brown dwarf companion DH~Tau~B
associated with the classical T~Tauri star DH Tau.  Near-infrared
coronagraphic observations with CIAO on the Subaru Telescope have
revealed DH~Tau~B with $H=$~15~mag located at 2\,\farcs3 (330 AU) away
from the primary DH~Tau~A.  Comparing its position with a {\it
Hubble Space Telescope} archive image, we confirmed that DH~Tau~A and
B share the common proper motion, suggesting that they are physically
associated with each other.  The near-infrared color of DH~Tau~B is
consistent with those of young stellar objects.  The near-infrared
spectra of DH~Tau~B show deep water absorption bands, a strong K~I
absorption line, and a moderate Na~I absorption line.  We derived its
effective temperature and surface gravity of $T_{\rm eff} =$~2700 --
2800~K and $\log g =$~4.0--4.5, respectively, by comparing the
observed spectra with synthesized spectra of low-mass objects.  The
location of DH~Tau~B on the HR diagram gives its mass of 30 --
50~M$_{\rm Jupiter}$.
\end{abstract}


\section{INTRODUCTION}

Brown dwarfs are objects with their masses less than 0.08~M$_{\sun}$.
Unable to sustain stable hydrogen burning in their cores, they
contract monotonously as they release gravitational energy into
radiation.  Their luminosities are small; even a 0.07~M$_{\sun}$ brown
dwarf has a luminosity of only 10$^{-4}$~L$_{\sun}$ at its age of
10$^{9}$~yr (Burrows et al. 1997;\markcite{Burrows} Baraffe et
al. 2003\markcite{Baraffe}).  Their faintness has prevented their
discovery despite an early prediction of their existence (Hayashi \&
Nakano 1963\markcite{Hayashi}).  Since Gl229B was first identified as
a bona fide brown dwarf (Nakajima et al. 1995\markcite{Nakajima};
Oppenheimer et al. 1995\markcite{Oppenheimer}), more than 300 brown
dwarfs have been discovered by photometric and spectroscopic
observations including Doppler shift measurements.  While more than
half of stars are found in binary systems, only $\sim 20$ brown dwarfs out of
300 have so far been identified as companions to stars.

The deficiency of companion brown dwarfs led some researchers to favor
gravitational collapse of molecular cloud cores as their formation
mechanism, similar to that of stars (Bonnell \& Bastien
1992\markcite{Bonnell}; Bate 2000\markcite{Bate}).  It has, on the
other hand, also been proposed that brown dwarfs may form from
circumstellar disks as companions to stars (Rice et
al. 2003\markcite{Rice}), similar to the possible mechanism of giant
planet formation initiated by gravitational instability of a disk.
Successive dynamical evolution of such a star and brown dwarf system
with other stars in a newly formed cluster may eject the brown dwarf
from the system (Reipurth \& Clarke 2001\markcite{Reipurth}).  It is
thus important to find young brown dwarf companions to study their
formation mechanism.

Although young brown dwarfs are mostly found in star forming regions
located far away when compared with nearby stars, they are bright and
relatively easy to be detected; a 0.03~M$_{\sun}$ brown dwarf is
10$^{-2}$~L$_{\sun}$ at its age of 10$^{6}$~yr (Burrows et
al. 1997;\markcite{Burrows} Baraffe et al. 2003\markcite{Baraffe}).
Near-infrared imaging with high sensitivity indeed detected many young
brown dwarfs in star forming regions (e.g. Oasa et
al. 1999\markcite{Oasa}; Zapatero-Osorio et al. 2002\markcite{Zapatero};
Allen et al. 2002\markcite{Allen}; Mainzer \& McLean 2003\markcite{Mainzer};
Luhman et al. 2003\markcite{Luhman}). Only a small number of
brown dwarf companions, however,
has so far been found around young stars (CoD$-33^{\circ}7795$ B,
Lowrance et al. 1999;\markcite{Lowrance99} GG Tau Bb, White et
al. 1999;\markcite{White99} HR 7329 B, Lowrance et
al. 2000\markcite{Lowrance00}), as is similar to the case of
companions to main sequence stars.
Very recently, Chauvin et al. (2004)\markcite{Chauvin} 
announced the discovery of a 
planetary-mass companion candidate to a young brown dwarf.

In this article we report the detection of a faint brown dwarf
companion to DH~Tau with CIAO (Coronagraphic Imager with Adaptive
Optics; Tamura et al. 2000\markcite{Tamura}) mounted on the Subaru
Telescope.  DH~Tau is a classical T Tauri star associated with the
Taurus molecular cloud and has a visual extinction of 0.0--1.5~mag
(Strom et al. 1989;\markcite{Strom} White \& Ghez
2001\markcite{White01}).  It has a mass and age of
0.25--0.5~M$_{\sun}$ and 10$^{5}$--4$\times 10^{6}$~yr, respectively
(Hartigan et al. 1994;\markcite{Hartigan} White \& Ghez
2001\markcite{White01}), where the uncertainties arise mainly from that of
bolometric luminosity.  The companion is located 330 AU from the
primary and has the effective temperature, mass and age of 2700 --
2800~K, 0.03 -- 0.05~M$_{\sun}$ and $3\times10^{6}$ -- $10^{7}$~yr,
respectively.

We are conducting a near-infrared coronagraphic survey for proto-planetary 
disks and faint companions around T Tauri stars. DH Tau was initially 
selected as a target for a proto-planetary disk. Its polarization 
angle is perpendicular to the nearby magnetic field direction, implying the
existence of a circumstellar disk (Tamura \& Sato 1989\markcite{Tamura89}). 
Other target 
samples consist mainly of objects with circumstellar disks detected 
in the millimeter wavelengths (Kitamura et al. 2003\markcite{Kitamura}). 
By near-infrared
coronagraphic observations, a dozen of faint point sources are detected
in the vicinity of the T Tauri stars. However, it is not yet confirmed 
by spectroscopic observations and proper motion measurements
whether the faint sources are associated with the T Tauri stars or 
are background objects.  The nature of these faint sources will be 
investigated in the future.


\section{OBSERVATIONS AND DATA REDUCTION}

The coronagraphic observations were carried out at $H$-band on 2002
November 23 and at $J$, $H$ and $K$-bands on 2004 January 8 with CIAO,
which was equipped with a 1024$\times$1024 InSb Alladin~II detector
with a spatial scale of 0\farcs0213 pixel$^{-1}$.  The spatial
resolution provided by the adaptive optics system was 0\farcs1 (FWHM)
under the natural seeing of $\sim$0\farcs5.  An occulting mask made of
chrome on a sapphire substrate had a diameter of 0\farcs6, within
which the transmittance was a few tenths of a percent.  This allowed
us to measure the accurate position of the central object.  We used a
traditional circular Lyot stop with its diameter 80~\% of the pupil.

For taking images at each band, we adjusted the telescope pointing
finely so that the DH~Tau primary (hereafter called DH~Tau~A) was
placed at the center of the occulting mask.  Three exposures of 10~sec
each were coadded into one frame.
After 6 frames of observations, both the telescope and the occulting mask
were dithered by $\sim1"$. DH Tau A was again placed at the center of the
occulting mask, then additional 18 frames were taken. The total exposure time is
720 sec in each band.
We observed SAO 76560 immediately after DH~Tau as a reference
star of the point spread function (PSF) under the same configuration.
We also observed FS 111 for photometric calibration.  Dark frames and
dome flats with incandescent lamps were taken at the end of the
night.

The Image Reduction and Analysis Facility (IRAF) was used for data
reduction.  A dark frame was subtracted from each object frame, then
each object frame was divided by the dome-flat.  Hot and bad pixels
were removed from the frame.  The peak position of the PSF moved
slightly on the detector during the observations.  This was caused by
the difference in the atmospheric distortion between the infrared
wavelength at which the images were taken and the optical wavelength
at which the wavefront was sensored.  We measured the peak positions
of reference star PSF images with the RADPROFILE task and shifted them
to have the same peak position.  Since the Subaru Telescope is an
alt-azimuth telescope and we used an instrument rotator, the position
angle of the spider pattern changed with time.  The reference star
frames were hence rotated so that the position angle of the spider
matched that of each object frame, and then combined.  The combined
reference star PSF was shifted so that its peak position coincided
with the DH~Tau~A peak, which had been measured with the RADPROFILE
task in each frame.  The reference star PSF was then renormalized so
that its wing intensity level became the same as that of DH~Tau~A at
$\sim1\arcsec$ away from its peak.  The halo of DH~Tau~A was
suppressed in each frame after the reference star PSF was subtracted,
which allowed us to detect an object (hereafter called DH~Tau~B) that
was located at 2\,\farcs34 southeast of DH~Tau~A and was $\sim6$~mag
fainter.  Finally, all the 24 object frames were combined into one
image.

The $K$-band spectroscopy of DH~Tau~B was made on 2003 November 8 with
CISCO mounted on the Subaru Telescope.  We used a grism with a
spectral resolution of $R~(=\lambda/\Delta\lambda) \sim440$ at
2.2~$\mu$m and a slit width of $0\farcs5$ under the typical seeing of
0\farcs5.  The adaptive optics system was not available for CISCO.  In
order to minimize the contamination of light from DH~Tau~A, we placed
the slit on DH~Tau~B to be perpendicular to the line connecting
DH~Tau~A and B.  Using the $K$-band images taken just
before the spectroscopic observations, 
we estimate that contaminated flux from the
primary in the spectra is $\sim10~\%$ of the flux of the secondary.
DH~Tau~B was observed at 2 slit positions with a $10\arcsec$ dithering
to obtain sky data simultaneously.  Two exposures of 300~sec each were
made at each slit position.  SAO~76860 (A0V) was used as a standard
star for spectral calibration.  The $JH$-band spectra were observed in
the similar manner on 2004 January 9 with $R \sim300$ and the total
integration time of 60~min.

We used IRAF for spectral data reduction again.  Each dithered pair of
object frames were subtracted from each other and divided by a flat
field frame.  We then geometrically transformed the frames to remove
the curvature of the slit image caused by the grating.  Wavelengths
were calibrated using the OH lines in the sky portion of the spectrum
with the uncertainty of 10~\AA~rms.  Individual spectra were extracted
from the transformed images using the APALL task.  The areas of the
spectral images where the intensity of DH~Tau~B was more than 20\% of
its peak intensity at each wavelength were summed into an one-dimensional
spectrum.  The spectrum was then normalized and combined to produce the
final spectrum.  After the hydrogen absorption lines in the standard
star spectrum were interpolated and removed, the object spectrum was
divided by the standard star spectrum and was multiplied by a
blackbody spectrum representing the standard star.  The $JH$- and $K$-band
spectra were scaled to the photometric magnitudes.

\section{RESULTS AND DISCUSSION}

\subsection{A Physical Companion}

\begin{figure}
\plotone{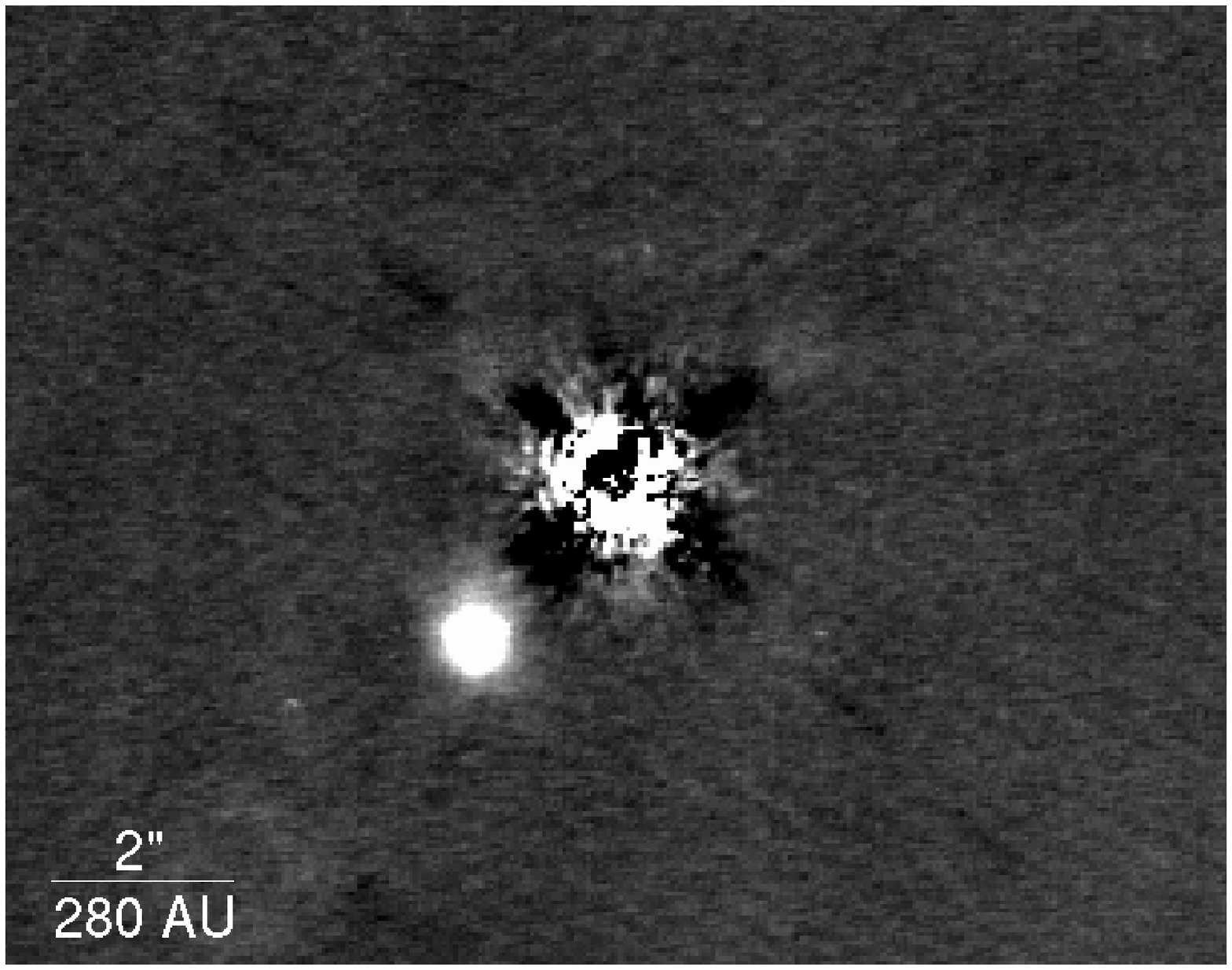}
\caption{$K$-band coronagraphic image of DH Tau. 
North is up and east is to the left.
The primary star (DH~Tau~A) is located at the center of the image but is blocked by the mask.
The companion (DH~Tau~B) is detected at $\sim 2\farcs34$ south-east of DH~Tau~A.
\label{dhres}}
\end{figure}

The $K$-band coronagraphic image of DH~Tau is presented in
Figure~\ref{dhres}. DH~Tau~A, occulted by the mask, is located at the
center, where the bright speckles are the residual halo of PSF
subtraction.  We do not detect any diffuse circumstellar structure.  
While near- and mid-infrared excesses and millimeter flux, possibly from 
a circumstellar structure are reported (Meyer et al. 1997a; 
Dutrey et al. 1996), any spatial-resolved image of the circumstellar 
structure has not been presented. Even with the coronagraph, we do not expect 
any detection of a circumstellar structure with such a short integration time, 
unless it is highly-flared (Itoh et al. 1998)\markcite{Itoh98}.

A point source (DH~Tau~B) was detected at the southeast of DH~Tau~A.  We
consider from the following arguments that DH~Tau~B is most likely a
physical companion to DH~Tau~A.
The offset of DH~Tau~B with respect to A was ($\Delta\alpha,
\Delta\delta) = (+1\farcs512\pm0\farcs003, -1\farcs791\pm0\farcs003$)
on 2004 January 8, corresponding to the angular separation and
position angle of $2\farcs344\pm0\farcs003$ ($328.2\pm0.4$ AU) and
$139\arcdeg.83\pm0\arcdeg.06$, respectively.  These were
$2\farcs340\pm0\farcs006$ and $139\arcdeg.56\pm0\arcdeg.17$,
respectively, on 2002 November 23 ( $(\Delta\alpha, \Delta\delta) =
(+1\farcs518\pm0\farcs008, -1\farcs781\pm0\farcs008)$ ).  
The uncertainties are the standard deviation of the measurements of positions
in each frame (0.008" and 0.002" both in the $\alpha$ and $\delta$ directions, 
for the 2002 and 2004 observations respectively), and the uncertainties in
the pixel scale and in the orientation of the detector
\footnote{ The pixel scale and the orientation of the detector were determined
by observations of the Trapezium cluster (Simon et al. 1999)\markcite{Simon}.
The pixel scale for the 2002 observations is $0.02125\pm0.00003"$ pixel$^{-1}$.
Note that the pixel scale in Itoh et al. (2002)\markcite{Itoh02} is misprinted.
The detector of CIAO was replaced in 2003 Nov.
The pixel scale for the new detector is $0.02133\pm0.00002"$ pixel$^{-1}$.}.
Since any other object is not found in the small field of view of CIAO, and
DH Tau B has not been recorded in any published catalog, the offsets of
DH Tau B are derived only with the positions of DH Tau A and B.

\begin{figure}
\plotone{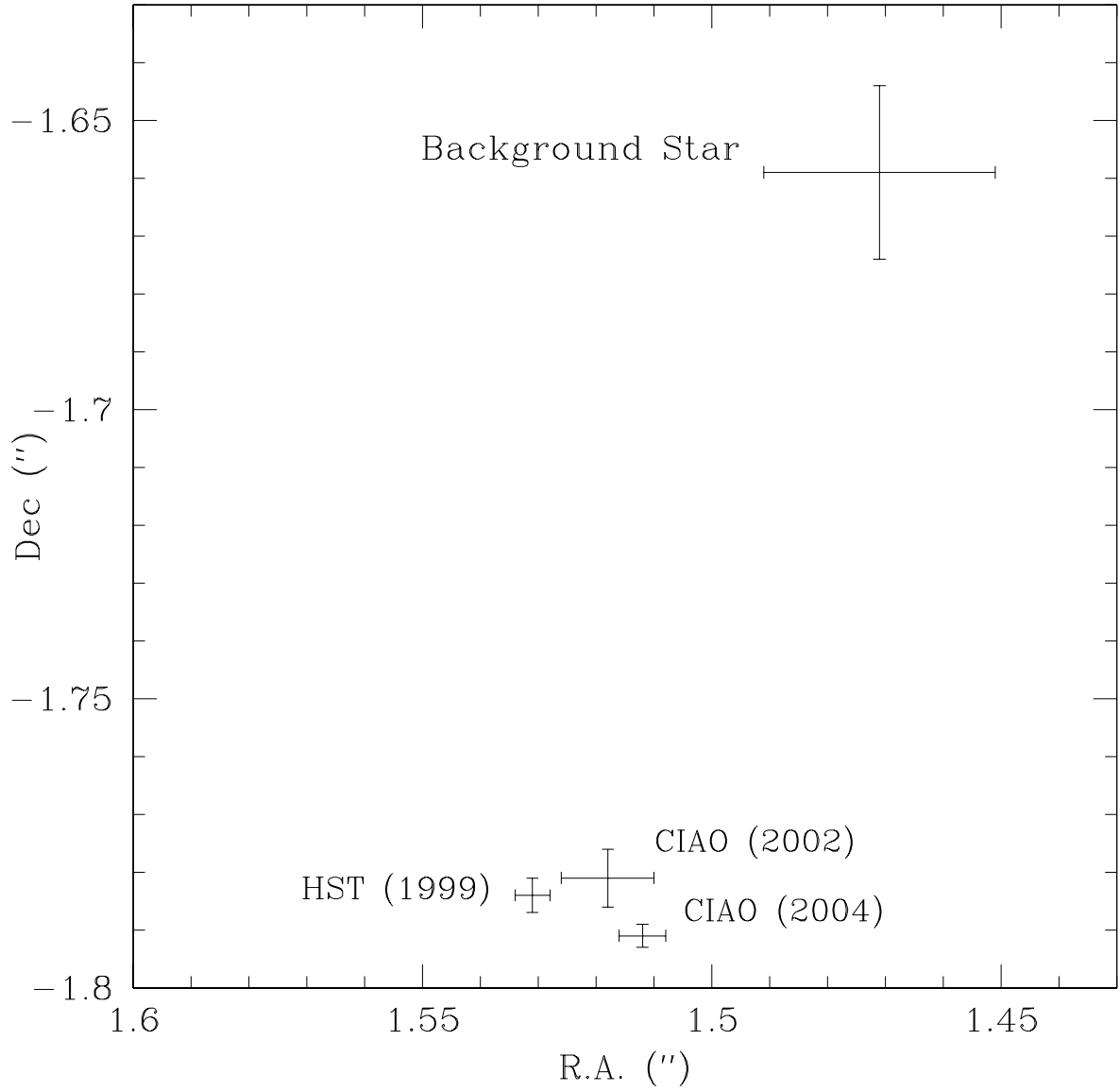}
\caption{Astrometric measurements of DH~Tau~B.  Offset positions are
shown with respect to DH~Tau~A.  The expected position of a distant
background star on 2004 January is shown by assuming that it was
located at the {\it HST} position on 1999 January.
\label{astrometry}}
\end{figure}

DH~Tau~B was
also detected in the {\it Hubble Space Telescope} archive data imaged
on 1999 January 17 with the offset of
($\Delta\alpha=+1\farcs531\pm0\farcs003$,
$\Delta\delta=-1\farcs784\pm0\farcs003$) or the separation and
position angle of $2\farcs351\pm0\farcs001$ and
$\theta=139\arcdeg.36\pm0\arcdeg.10$, respectively (Holtzman et al. 1995
\markcite{Holtzman} for the pixel scale).  The three
measurements give consistent offsets within 0\farcs02, comparable to
the pixel size of CIAO, in both $\alpha$ and $\delta$ directions,
suggesting that DH~Tau~A and B share a common proper motion in five
years.  Because DH~Tau~A has a proper motion of
$\delta_{\alpha}\sim+12\pm4$~mas/yr and
$\delta_{\delta}\sim-25\pm3$~mas/yr (Monet et
al. 2003;\markcite{Monet} Hanson et al. 2003;\markcite{Hanson}
Zacharias et al. 2003\markcite{Zacharias}) that could be easily
detected against a background or foreground star with CIAO in five
years of interval, DH~Tau~B should not be a background or foreground
star.  In fact, DH~Tau~B would have moved by ($-0\farcs060$,
$0\farcs125$) in five years if it had been a background star (see
Figure \ref{astrometry}).

DH~Tau~B is hence associated with the Taurus molecular cloud, where
new born stars share a similar proper motion of
$\delta_{\alpha}\sim+6\pm7$~mas/yr and
$\delta_{\delta}\sim-22\pm6$~mas/yr (Jones \& Herbig
1979\markcite{Jones}).  Two physically unrelated stars associated with
the Taurus molecular cloud have a large chance to have similar proper
motions that we cannot discern in five years.  There is, however, a
very small chance that an object with $J=15$ happens to be located
within $2\farcs3$ from DH Tau A.  We expect 20 such objects in 1
arcdegree square, by extrapolating the luminosity function of low-mass
objects associated with the Taurus molecular cloud (Itoh et
al. 1996\markcite{Itoh}).  If such objects are randomly distributed,
chance probability that an isolated object is located at
$2\farcs3$ from DH Tau A is estimated as small as $10^{-5}$.  We
therefore conclude that DH~Tau~B is most likely a physical companion
to DH~Tau~A.

\subsection{The Near-Infrared Spectrum}

\begin{figure}
\plotone{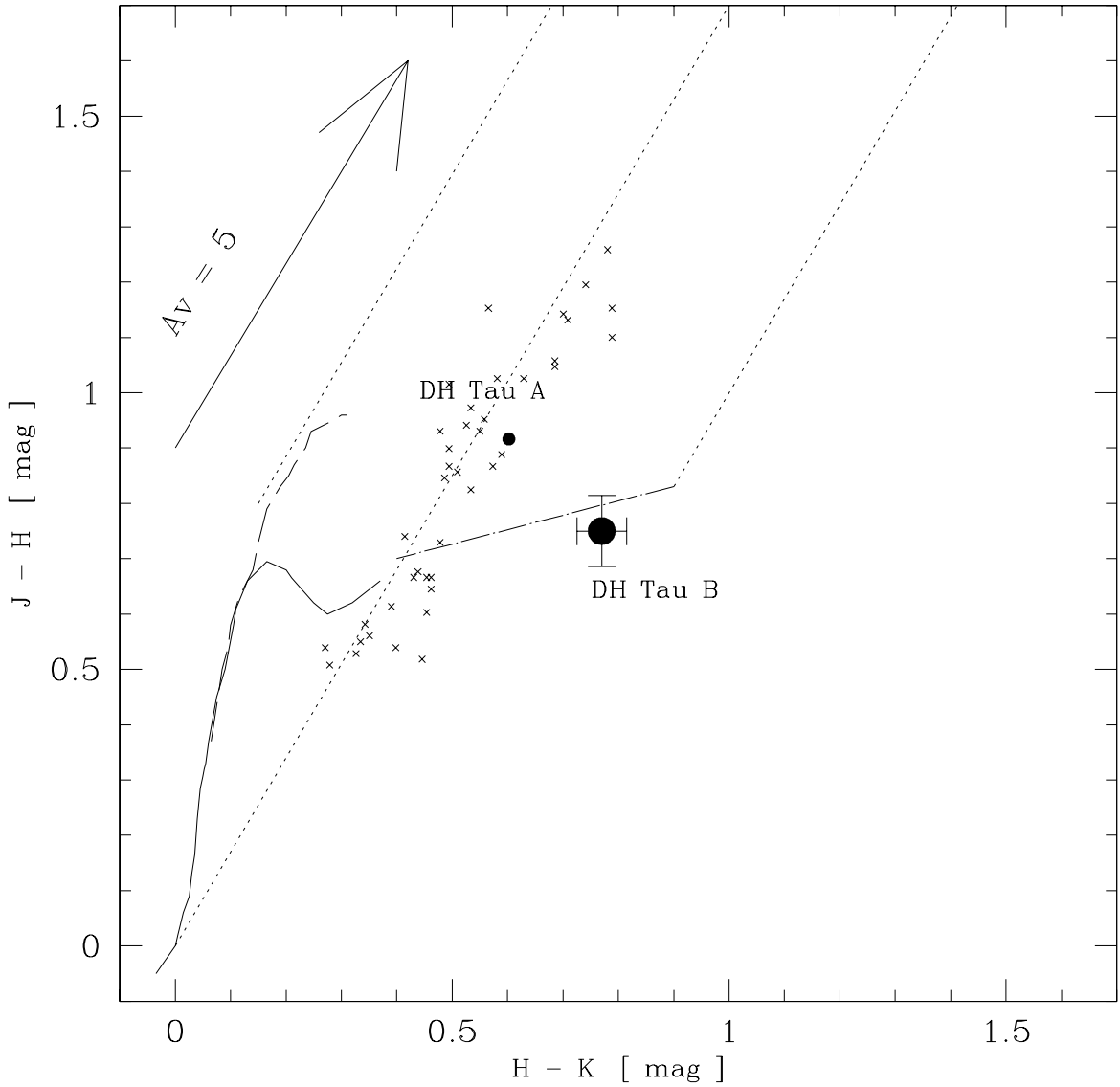}
\caption{The near-infrared colors of DH~Tau~A and B.  Also plotted are
the near-infrared colors of main-sequence stars (solid line; Bessell
\& Brett 1988), giants (dashed line; Koornneef 1983), classical T
Tauri stars (dot-dashed line; Meyer et al. 1997b), and L-type brown
dwarfs (crosses; Leggett et al. 2002).  All colors are transformed
into the CIT system.
\label{cc}}
\end{figure}

Photometry of DH~Tau~B gives its near-infrared magnitudes of
$15.71\pm0.05$, $14.96\pm0.04$ and $14.19\pm0.02$ at the $J$-, $H$-
and $K$-bands, respectively.  Its colors, together with those of
DH~Tau~A, are plotted on the two color diagram in Figure~$\ref{cc}$.
Both DH~Tau~A and B have the colors consistent with T~Tauri stars.
Note that the redder color of DH Tau B than DH Tau A does not explicitly
indicate larger infrared excess of DH Tau B than DH Tau A.
Because DH Tau B has a lower effective temperature as derived below
(2700 K -- 2800 K), than DH Tau A (3580 K; Hartigan et al. 1994),
photospheric color of DH Tau B is expected to be redder than that of DH Tau A.
Assuming DH Tau A and B have the same photospheric colors as
early- and late-M dwarfs, respectively, infrared excesses of
both objects are comparable.

\begin{figure}
\plotone{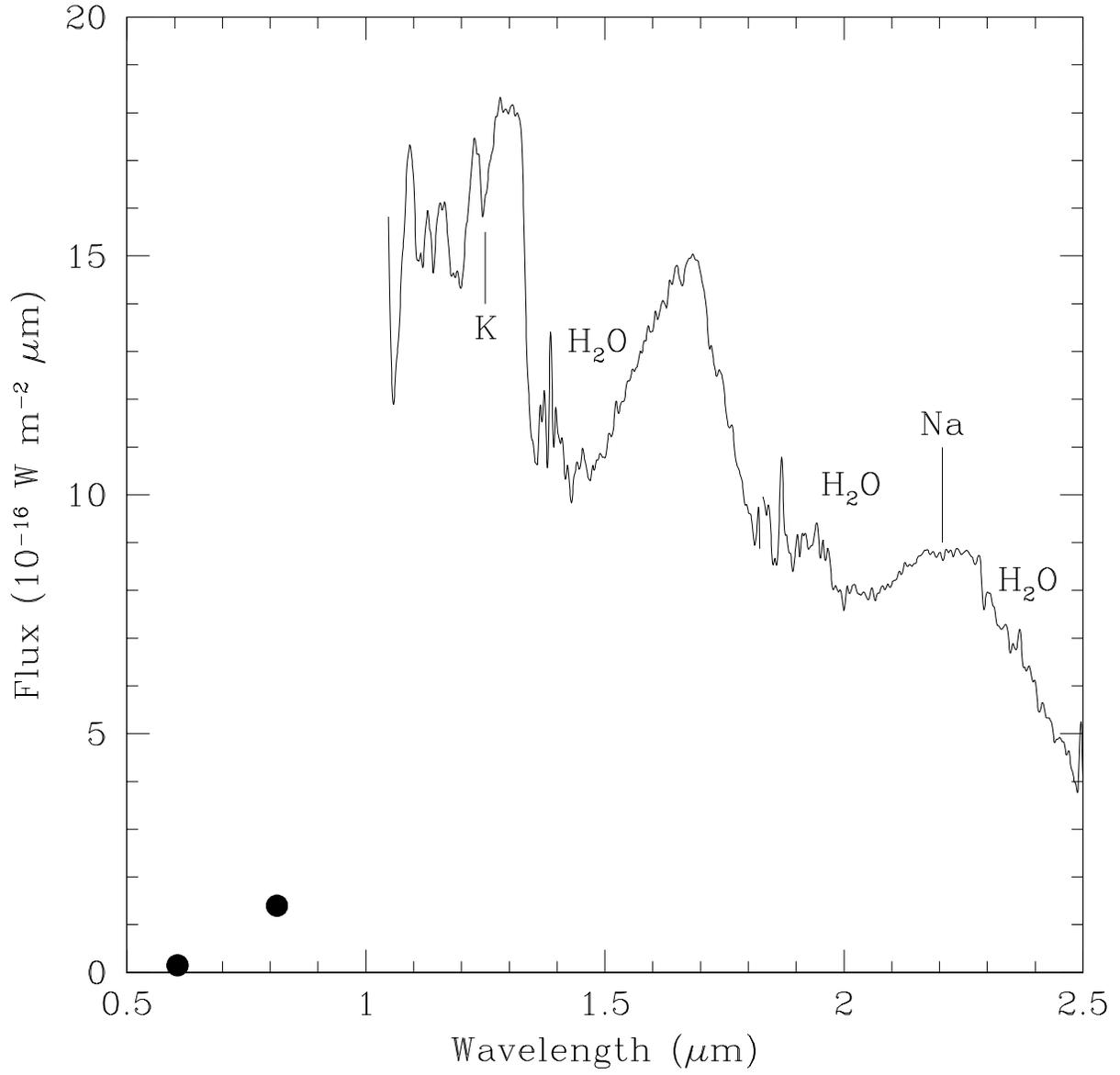}
\caption{The near-infrared spectra of DH~Tau~B.
Optical broad-band fluxes were measured on the {\it HST} archive data.
\label{specraw}}
\end{figure}

The near-infrared spectra of DH~Tau~B are shown in
Figure~$\ref{specraw}$, where prominent water absorption bands are
seen around 1.4~$\micron$ and 1.9~$\micron$.  
These absorption bands are also prominent in the $H$-band spectrum of 
the giant planet candidate around a young brown dwarf (Chauvin et al. 2004
\markcite{Chauvin}).
The absorption lines of
K~I and Na~I were detected at 1.25~$\micron$ and 2.21~$\micron$,
respectively.  Their equivalent widths are measured to be
$6.8\pm2.8\rm\AA$ and $1.8\pm1.4\rm\AA$ for the K~I and Na~I lines,
respectively, using the spectro-photometric bands defined by Gorlova
et al. (2003)\markcite{Gorlova}.  These characteristics suggest a low
temperature of DH~Tau~B.

\begin{figure}
\plotone{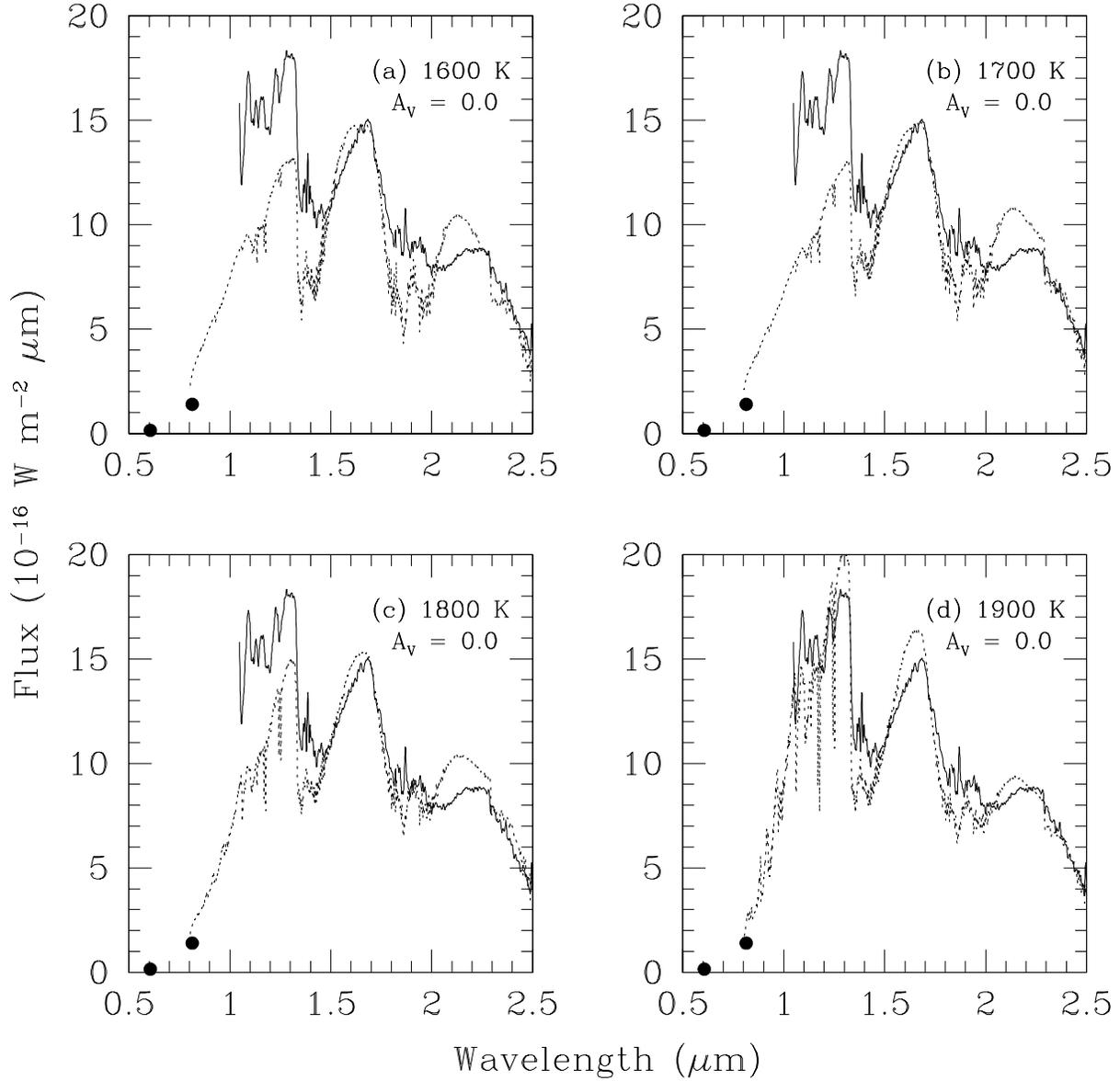}
\caption{The near-infrared spectra of DH~Tau~B (solid line) are shown
in each panel together with the best fit synthesized spectrum with
$\log g =5.5$ (dotted line) for the effective temperature ranging from
1600~K to 2900~K.  The assumed visual extinction is also shown in each
panel.
\label{tsuji5.5}}
\end{figure}

We compare the observed spectra with model calculations in
Figures~$\ref{tsuji5.5}a$--$l$ in order to estimate the spectral type
of DH~Tau~B, i.e., the effective temperature ($T_{\rm eff}$) and
surface gravity ($\log g$).  We used the model spectra from the data
provided by Tsuji et al. (2004)\markcite{Tsuji} by tentatively
assuming that DH~Tau~B is a dwarf ($\log g =$~5.5), although the
overall spectral shape has a strong dependence on $\log g$ at $T_{\rm
eff} <$~2500~K, which will be discussed later.  We also employed the
extinction of $A_{\rm V} =$~0 -- 4.4~mag in order to accommodate the
synthesized spectra to the observed near-infrared spectra and the
optical fluxes measured on the {\it HST} archive data.  Continuum
veiling due to the circumstellar disk is assumed only in the $K$-band,
with wavelength dependence as $F_{\lambda} \propto \lambda^{\alpha}$
($\alpha>0$).  The observed spectra were fitted by a synthesized
spectrum at each $T_{\rm eff}$ by minimizing the total residual at the
three wavelength ranges of 1.45 -- 1.60~$\mu$m, 1.70 -- 1.8~$\mu$m and
2.0 -- 2.45~$\mu$m.

Figures~$\ref{tsuji5.5}a$--$l$ show that the observed spectra are
overall well fitted by a model spectrum either with $T_{\rm
eff}=$~1900~K (Figure~$\ref{tsuji5.5}d$) or 2700 -- 2800~K
(Figures~$\ref{tsuji5.5}j$ and $k$).  Below 1800~K
(Figures~$\ref{tsuji5.5}a$--$c$), the calculated $J$-band fluxes are
significantly smaller than the observed flux.  The water absorption
bands of the models become too deep for $T_{\rm eff}$ between 2000~K
and 2600~K (Figure~$\ref{tsuji5.5}e$--$i$), resulting in the excess
$H$-band fluxes of model spectra.  The $J$-band flux of the model
again becomes too small at 2900~K.

\begin{figure}
\plotone{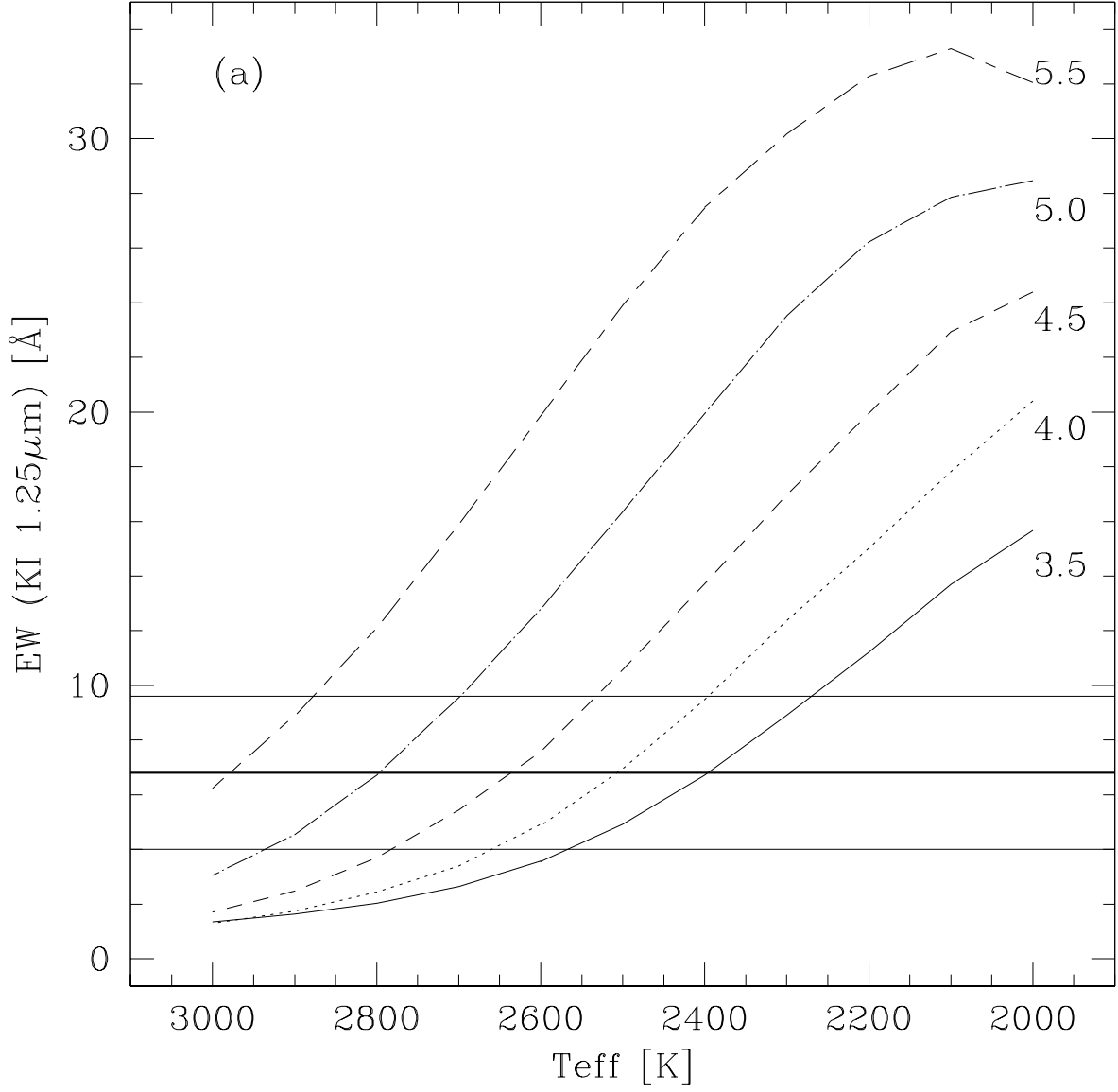}
\caption{(a) The equivalent width of the K~I line is shown as a
function of the effective temperature with each curve representing a
different value of the surface gravity $\log g$ given in its right.
The thick and thin lines are the observed equivalent width and its
error.  (b) Same as (a) but for the Na~I line.
\label{k}}
\end{figure}
\begin{figure}
\plotone{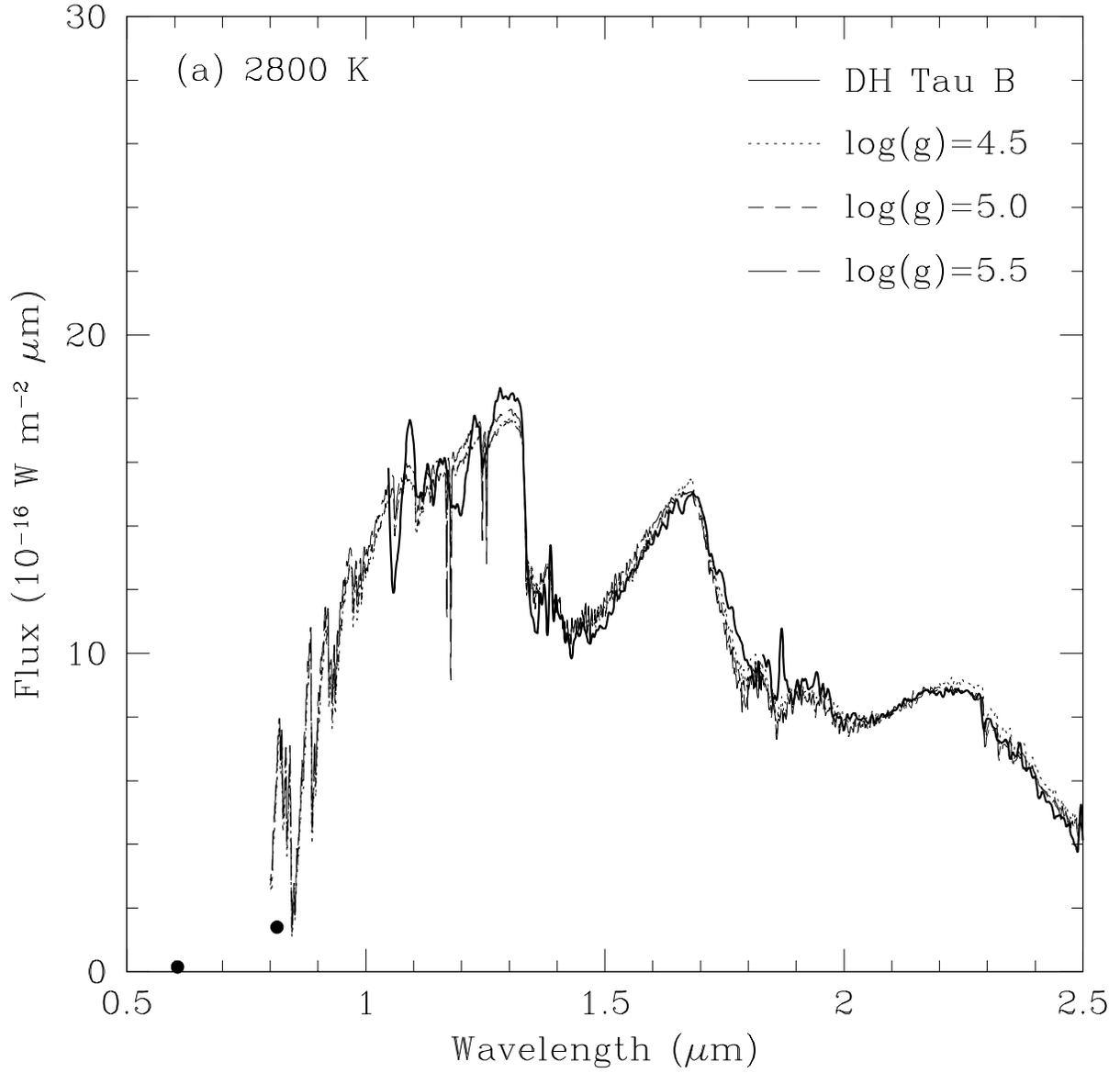}
\caption{(a) The near-infrared spectra of DH~Tau~B are compared with
the synthesized spectra with three different values of the surface
gravity for $T_{\rm eff}=2800$~K.  (b) Same as (a) but for $T_{\rm
eff}=1900$~K.
\label{tsuji2800}}
\end{figure}

In order to eliminate the temperature duality, we will use the
equivalent width (EW) of the K~I 1.25~$\mu$ m line.  Although the
overall shape of the observed spectra is fitted well in
Figures~$\ref{tsuji5.5}d$, $j$ and $k$, the observed strength of the
K~I line is always smaller than those of the model spectra.  We
calculated from the data provided by Tsuji et
al. (2004)\markcite{Tsuji} the EW of K~I as a function of $T_{\rm
eff}$ with various realistic values of $\log g$.  The results are
presented in Figure~$\ref{k}a$, showing that the EW is always larger
than the observed value for dwarfs ($\log g =$~5.5) for $T_{\rm eff}$
lower than 2900~K.  The observed EW is, however, accommodated with the
smaller values of $\log g=$~4.0 -- 5.0 if $T_{\rm eff}$ is in the
range between 2400~K and 2900~K.  None of the realistic values of
$\log g$ can account for the EW if $T_{\rm eff}$ is less than 2200~K.
We should therefore choose the higher $T_{\rm eff}$ solution for
DH~Tau~B, but its surface gravity should be smaller than that of a
dwarf.  The smaller surface gravity does not significantly alter the
overall shape of model spectra for the high $T_{\rm eff}$ solution as
shown in Figure~$\ref{tsuji2800}a$, while it gives unacceptable change
for the low $T_{\rm eff}$ solution as shown in Figure~$\ref{tsuji2800}b$.
McGovern et al. (2004)\markcite{McGovern}
also claimed that the weakness of KI lines is an indicative
of low surface gravity for an object.

The EW of the Na~I 2.21~$\mu$m line may give further constraints.
Figures~$\ref{k}b$ shows the dependence of its EW on $T_{\rm eff}$ and
$\log g$, again calculated from the data provided by Tsuji et
al. (2004).  The EW of Na~I increases with temperature especially when
$\log g$ is large.  The observed EW is accommodated with any
temperature between 2000~K and 3000~K if $\log g =$~3.5 -- 4.5, while
the model calculations predict significantly larger EWs if $\log g
\ge$~5.0 for $T_{\rm eff} >$~2600~K.  We therefore conclude that
DH~Tau~B has $T_{\rm eff} =$~2700 -- 2800~K and $\log g=$~4.0 -- 4.5.
The same values are derived when we use the model of Allard et
al. (2000).

\begin{figure}
\plotone{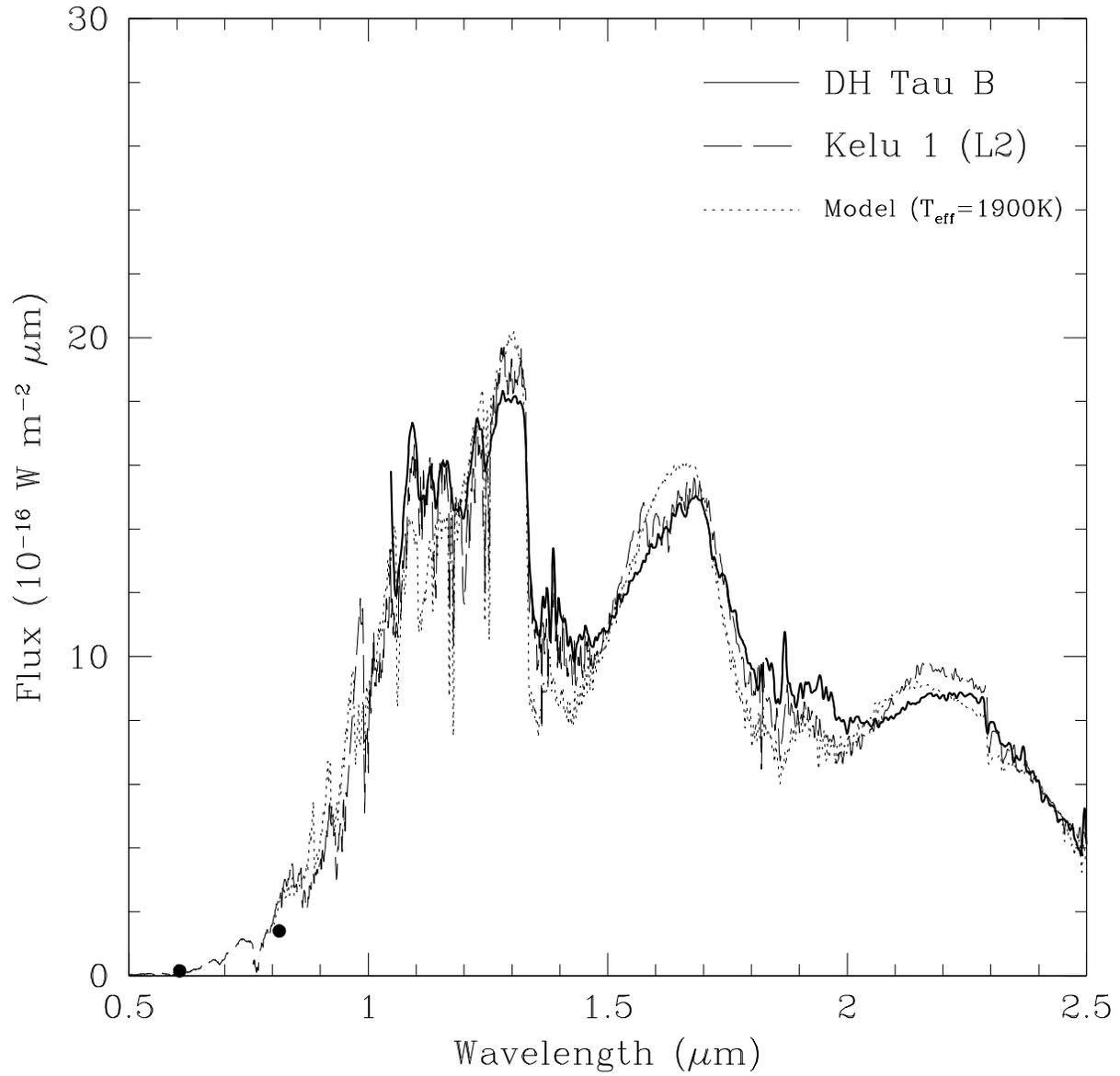}
\caption{The near-infrared spectra of DH~Tau~B are compared with that
of the L2 dwarf Kelu~1 (Leggett et al. 2001) and a synthesized
spectrum of $T_{\rm eff}=1900$~K and $\log (g)=5.5$.
\label{tsuji1900L3}}
\end{figure}

Near-infrared spectra of low mass young stellar objects have often
been compared to those of late type dwarfs ($\log g =$~5.5 -- 6.0).
Our observations have shown, however, that overall spectral shapes,
dominated by water absorption bands, are not sufficient to determine
the spectral type and effective temperature of these objects, but the
information of metallic lines is essential when the objects have
smaller surface gravity.  
Gravity effects in young brown dwarfs is firstly analized by Lucus et al.
(2001)\markcite{Lucas}.
An example of this is shown in
Figure~$\ref{tsuji1900L3}$, where the observed spectra of DH~Tau~B
($T_{\rm eff} =$~2700 -- 2800~K) are compared with that of Kelu~1 (L2,
$T_{\rm eff} =$~1900~K, Leggett et al. 2001)\markcite{Leggett} and a
model dwarf spectrum with $T_{\rm eff} =$~1900~K and $\log g$~=~5.5
taken from Tsuji et al. (2004)\markcite{Tsuji}.  The spectra of
DH~Tau~B are surprisingly similar to these low temperature dwarfs,
except that it shows only a shallow K~I absorption line.

\subsection{DH~Tau~B as a Brown Dwarf Companion}

\begin{figure}
\plotone{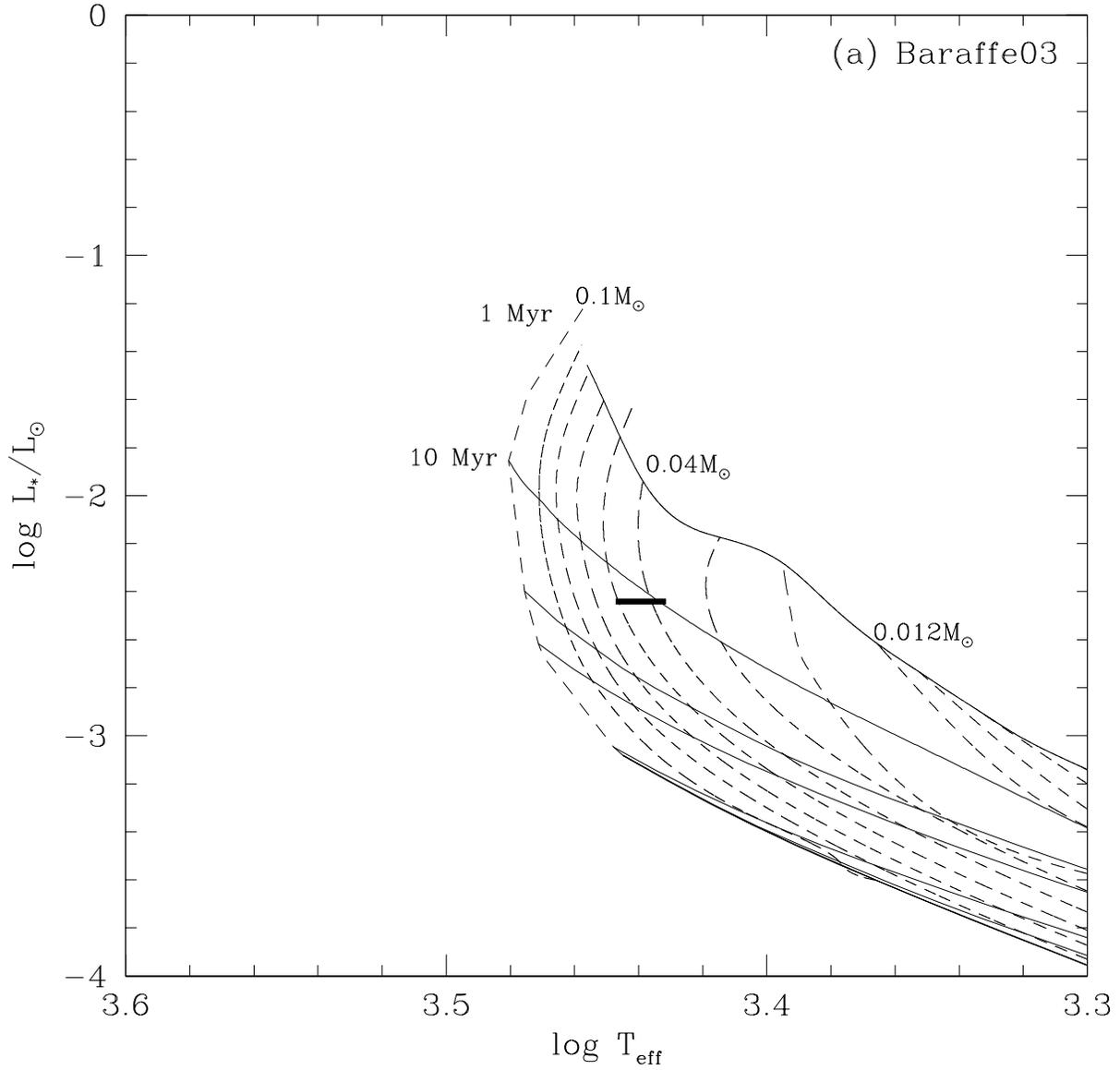}
\caption{(a) The location of DH~Tau~B on the evolutionary tracks of
Baraffe et al. (2003).  (b) same as (a) but on the evolutionary tracks
of D'Antona \& Mazzitelli (1997).
"H", "M", and "W" are the positions of DH Tau A (Hartigan et al. 1994;
Meyer et al. 1997a; White \& Ghez 2001).
\label{baraffe03}}
\end{figure}

The bolometric luminosity of DH~Tau~B is $\sim 10^{-2.44}~L_{\sun}$
derived from the synthesized spectra.  From this luminosity together
with the effective temperature, we estimated its mass and age of
0.04~--~0.05~M$_{\sun}$ and $10^{7}$~yr, respectively, from the HR
diagram of Baraffe et al. (2003\markcite{Baraffe}) as shown in
Figure~$\ref{baraffe03}a$.  If we use the result of D'Antona \&
Mazzitelli (1997\markcite{DM98}), the mass and ages are
0.03~--~0.04~M$_{\sun}$ and 3$\times10^{6}$--10$^{7}$, respectively
(Figure~$\ref{baraffe03}b$).  DH~Tau~B is thus a brown dwarf companion
to DH~Tau~A.  The age of DH~Tau~B appears to be larger than that of
$10^{5}$--$4\times10^{6}$~yr of DH~Tau~A.
Further discoveries of young brown dwarf companions, as well as precise
measurement of the bolometric luminosity of DH Tau A will solve
this mismatch in the age.

DH~Tau~A has another companion, DI~Tau, with a separation of
$16\arcsec$.  Holman \& Wiegert (1999)\markcite{Holman} investigated
the stability of such multiple systems and concluded that a less
massive tertiary has a stable orbit if its semi-major axis is less
than about a quarter of the separation of the binary.  DI~Tau is thus
located sufficiently far away to make the orbit of DH~Tau~B stable.
The orbital period of DH~Tau~B is estimated to be $\sim$6000~yr and
its orbital motion is $0\rlap{\arcdeg}\,.06$ yr$^{-1}$ if we assume a
pole-on geometry.  This would cause a position angle change of
DH~Tau~B with respect to A by $0\rlap{\arcdeg}\,.30$ between the 1999
{\it HST} and 2004 Subaru/CIAO observations.  The measured change of
the position angle was $0\rlap{\arcdeg}\,.47\pm0\rlap{\arcdeg}\,.12$
between the two epochs and may be within the consistent range.

Bonnell \& Bastien (1992)\markcite{Bonnell} and Bate
(2000)\markcite{Bate} proposed that companion brown dwarfs are formed
through the gravitational collapse of a molecular cloud, discussing
the deficiency of brown dwarf companions with small separations.
Brown dwarf companions with wider separations like the case of
DH~Tau~B are rather common according to this formation mechanism.
Another possible mechanism of companion brown dwarf formation is the
fragmentation of a circumstellar disk (Rice et
al. 2003)\markcite{Rice}.  Although this leads to the formation of a
number of low-mass companions, most of them, especially less massive
objects, escape from the system.  The remaining objects have large
semi-major axes ($a > 500$ AU) and large eccentricities.  The orbit
determination of DH~Tau~B will then provide a clue for distinguishing
the two mechanisms of companion brown dwarf formation.

\section{CONCLUSIONS}

We have detected a young brown dwarf companion to the classical
T~Tauri star DH~Tau using a coronagraphic camera CIAO on the Subaru
Telescope.  Their physical association was confirmed by the common
proper motion that the primary and secondary shared. 

(1) The companion brown dwarf has the $H$-band flux of 15~mag and is
    located at 2\farcs3 (330 AU) away from the primary.

(2) The near-infrared colors of the companion are consistent with
those of young stellar objects.  Its near-infrared spectra taken with
CISCO on the Subaru Telescope show deep water absorption bands and
strong K~I and moderate Na~I absorption lines, indicating its low
temperature.

(3) Comparison of the observed companion spectra with model spectra gave its
    effective temperature and surface gravity of $T_{\rm eff}=$~2700
    -- 2800 K $\log g=$~4.0 -- 4.5, respectively.

(4) We estimated the mass of 30 -- 50~M$_{Jupiter}$ for the companion
    from published evolutionary tracks of low-mass objects.

\acknowledgments

We are grateful for constructive and useful comments from an anonymous referee.
We also thank Kentaro Aoki and Sumiko Harasawa for their support
for making observations.  This work was supported by ``The 21st
Century COE Program: The Origin and Evolution of Planetary Systems" of
the Ministry of Education, Culture, Sports, Science and Technology
(MEXT).  Y. I. is supported by the Sumitomo Foundation, the Ito
Science Foundation, and a Grant-in-Aid for Scientific Research
No. 16740256 of the MEXT.
The HST data presented in this paper was obtained from 
the Multimission Archive at the Space Telescope Science Institute (MAST). 
STScI is operated by the Association of Universities for Research 
in Astronomy, Inc., under NASA contract NAS5-26555.
This work
is conducted as a part of the Observatory Project of "Subaru Disk and
Planet Searches".


\begin{figure}
\plotone{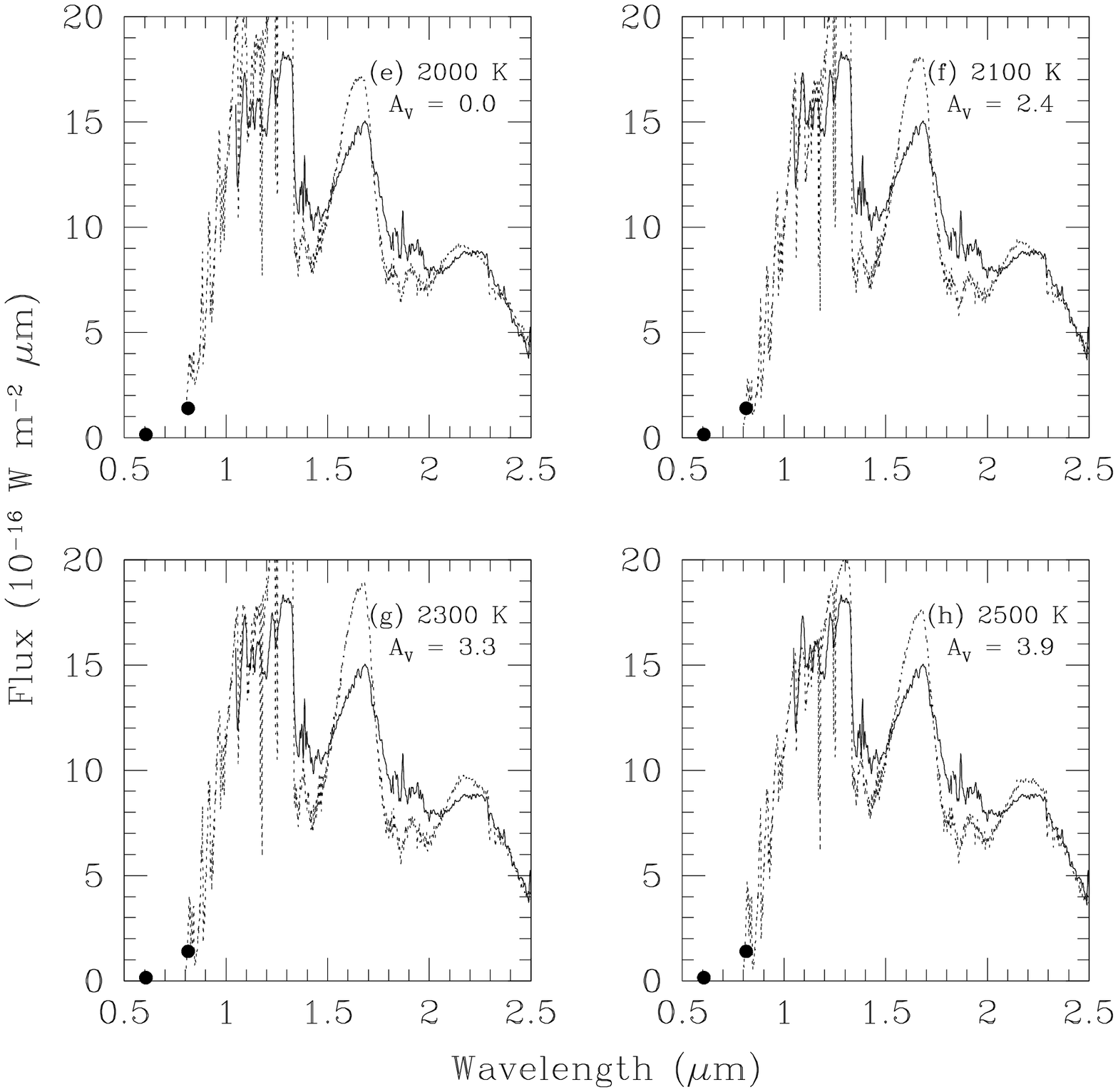}
\end{figure}

\begin{figure}
\plotone{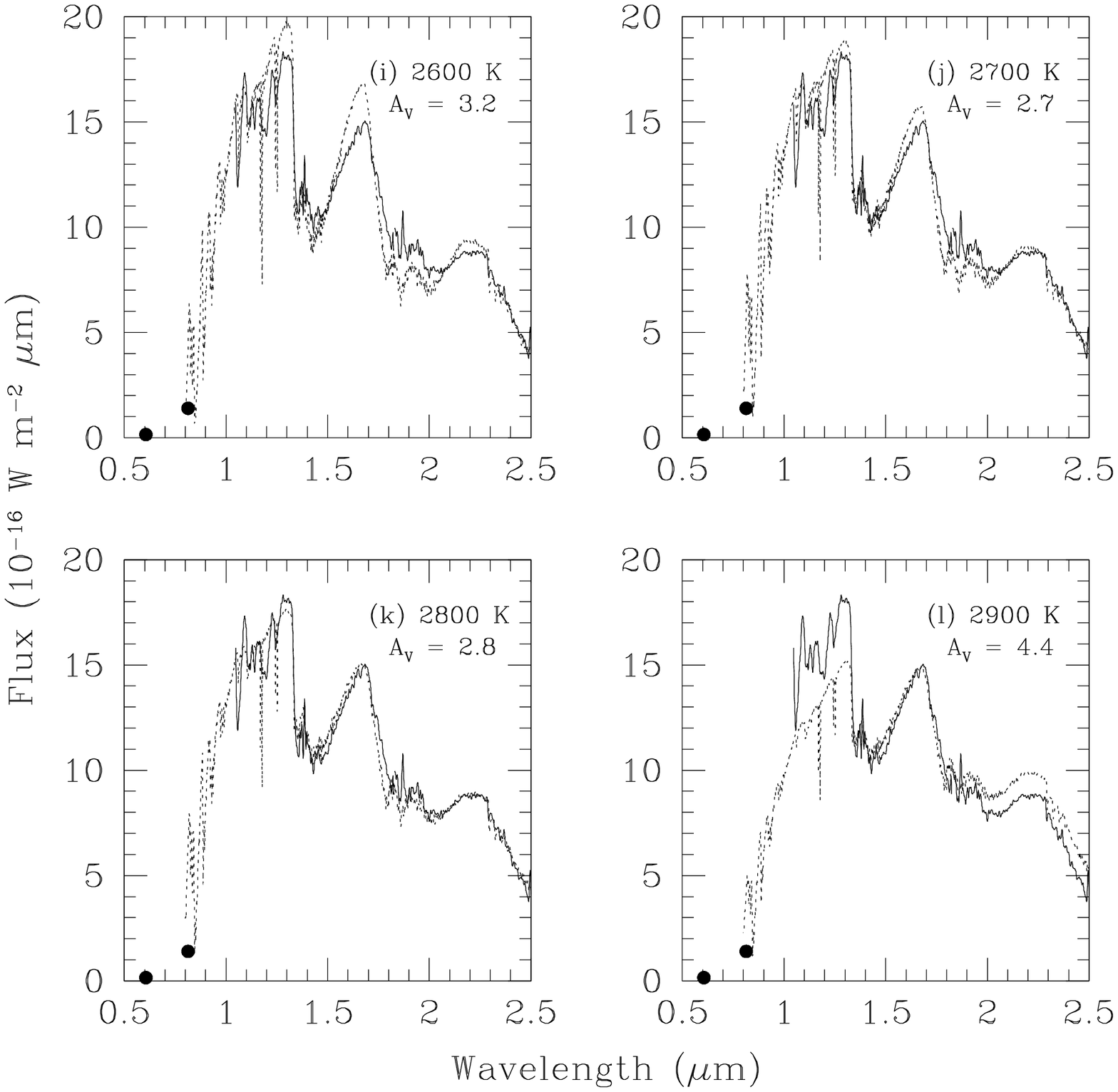}
\end{figure}

\begin{figure}
\plotone{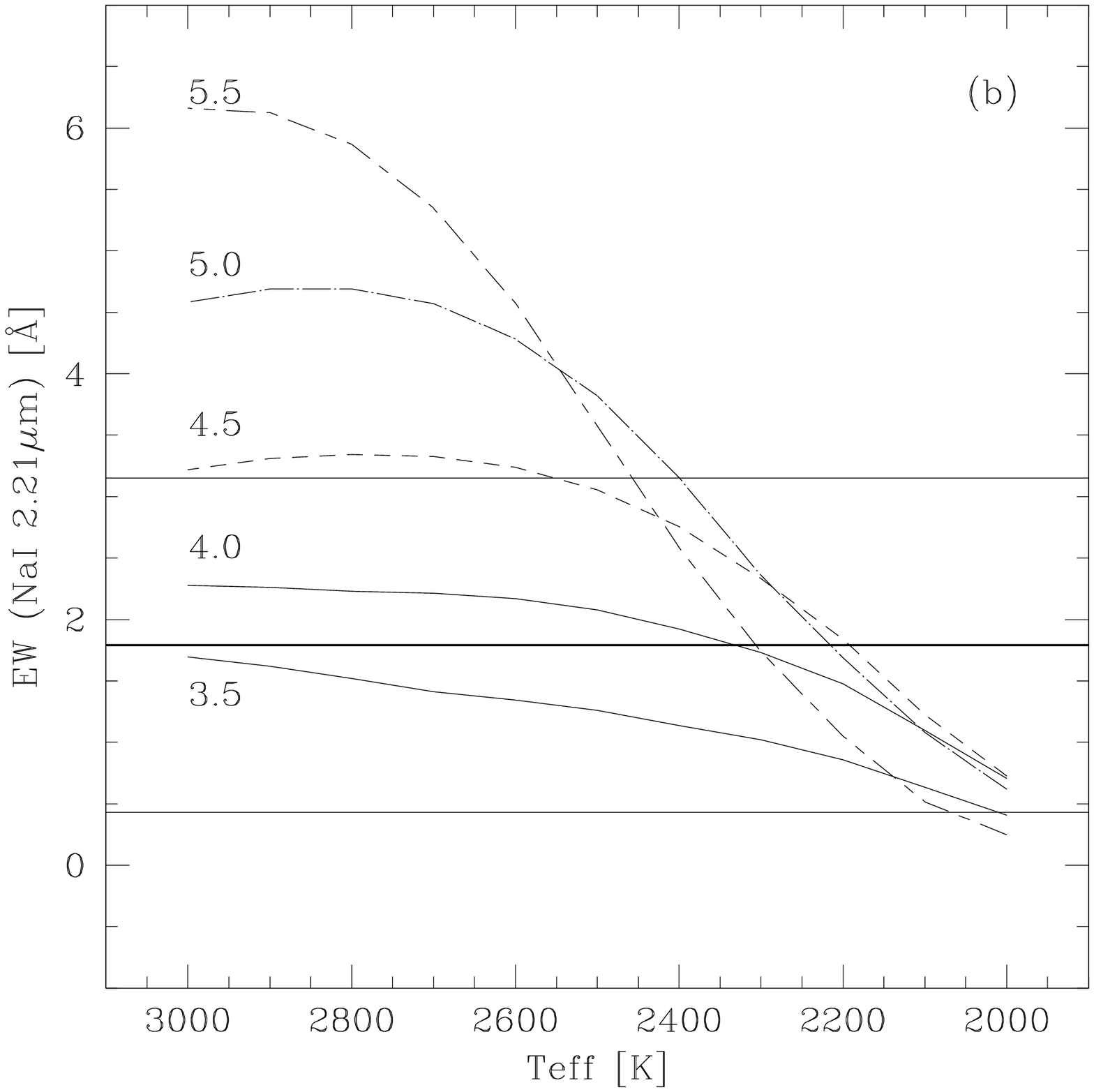}
\end{figure}


\begin{figure}
\plotone{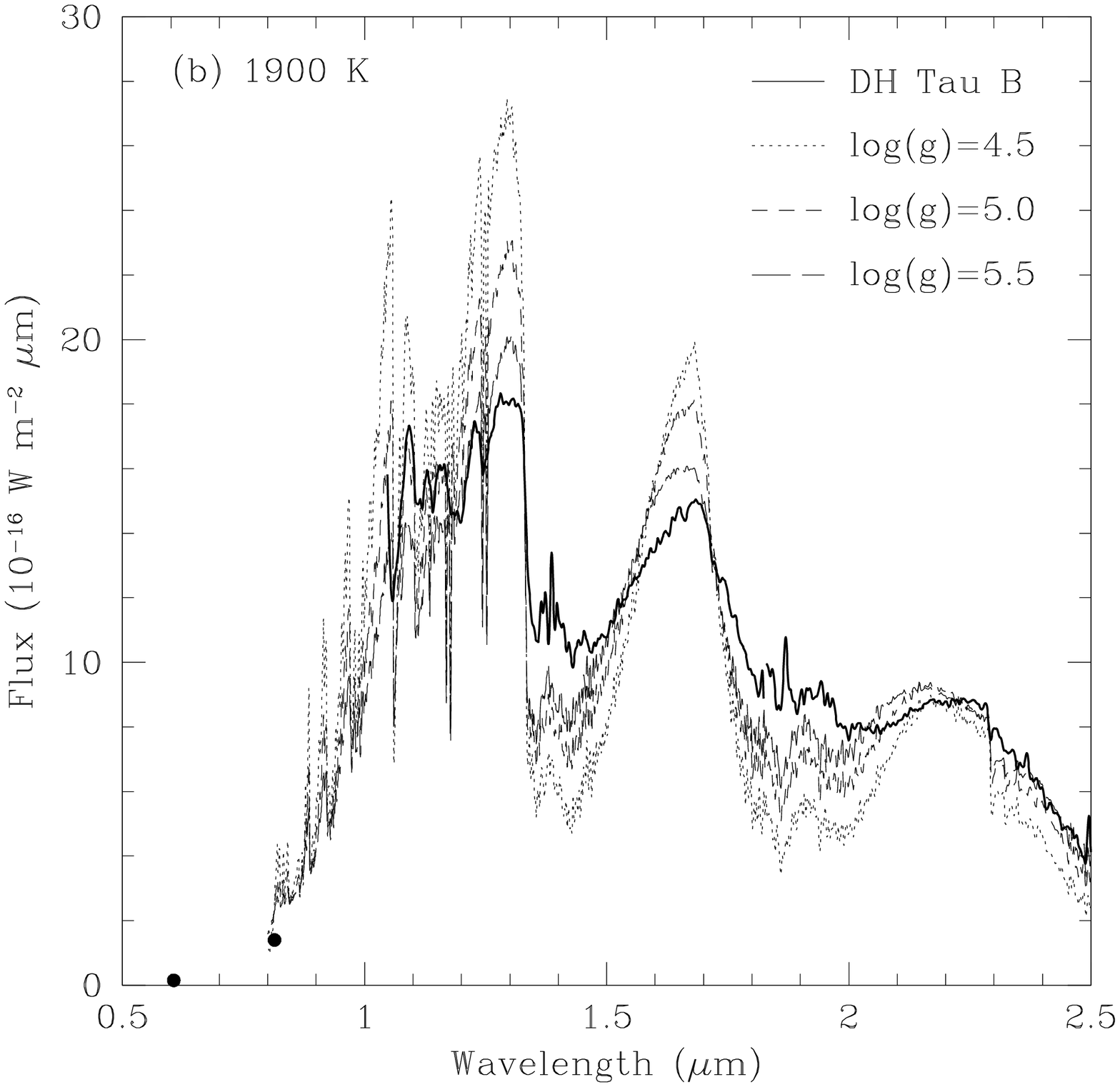}
\end{figure}

\begin{figure}
\plotone{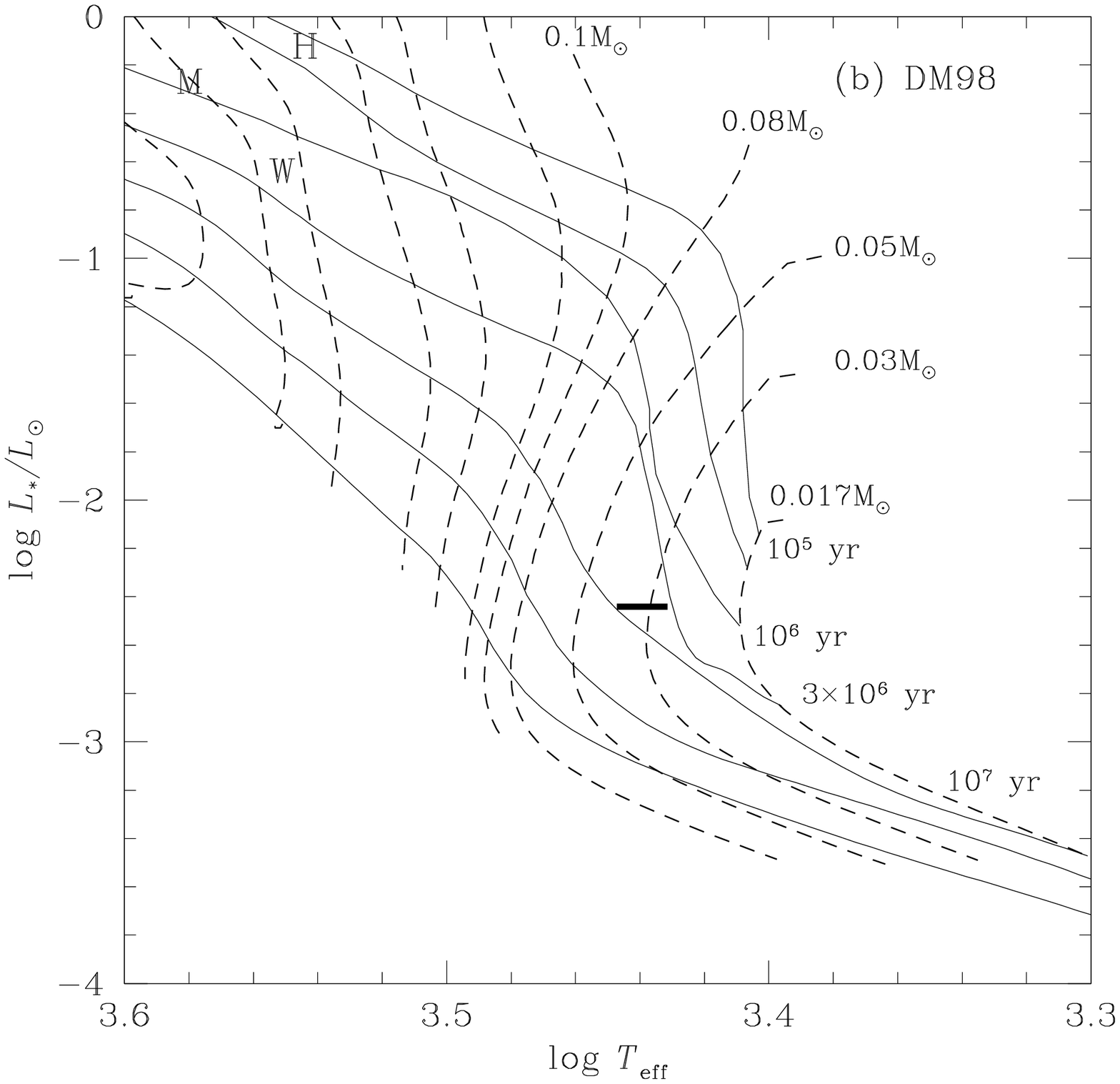}
\end{figure}

\clearpage

\begin{table}
\begin{center}
\caption{Observation log.\label{log}}
\begin{tabular}{lccc}
\tableline\tableline
date & band & exposure & total integ. time\\
\tableline
\multicolumn{4}{c}{Coronagraphy} \\
\tableline
2002 Nov. 23 & $H$ & 10 sec$\times$3 coadd$\times$24 frames & 720 sec\\
2004 Jan. 8  & $J$ & 10 sec$\times$3 coadd$\times$24 frames & 720 sec\\
             & $H$ & 10 sec$\times$3 coadd$\times$24 frames & 720 sec\\
             & $K$ & 10 sec$\times$3 coadd$\times$24 frames & 720 sec\\
\tableline
\multicolumn{4}{c}{Spectroscopy} \\
\tableline
2003 Nov. 8  & $K$  & 300 sec$\times$1 coadd$\times$4 frames & 1200 sec\\
2004 Jan. 9  & $JH$ & 300 sec$\times$1 coadd$\times$12 frames & 3600 sec\\
\tableline
\end{tabular}
\end{center}
\end{table}

\end{document}